\documentclass{PoS}

\newcommand{\Dslash}{D\hspace{-.23cm}/}
\newcommand{\partialslash}{\partial \hspace{-.23cm}/}
\newcommand{\doublearrowslash}{\stackrel{\leftrightarrow}{\partialslash}}
\newcommand{\doublearrowpartial}{\stackrel{\leftrightarrow}{\partial}}

\title{Nonlocal PNJL models and heavy hybrid stars}

\ShortTitle{Nonlocal PNJL models and heavy hybrid stars}

\author{\speaker{David Blaschke},
	David E. Alvarez Castillo\thanks{
	Instituto de F\'{\i}sica,  Universidad Aut\'onoma de San Luis
	Potos\'{\i}, San Luis Potos\'{\i}, S.L.P. 78290, M\'exico},\\
        Institute for Theoretical Physics, University of Wroc{\l}aw,
	Wroc{\l}aw, Poland\\
	Bogoliubov Laboratory for Theoretical Physics, JINR Dubna, Russia\\
        E-mail: \email{blaschke@ift.uni.wroc.pl, alvarez@ift.uni.wroc.pl}}

\author{Sanjin Beni\' c\\
        Physics Department, Faculty of Science,
        University of Zagreb, Zagreb 10000, Croatia\\
        E-mail: \email{sanjinb@phy.hr}}

\author{Gustavo Contrera\\
        CONICET, Rivadavia 1917, 1033 Buenos Aires, Argentina\\
        Gravitation, Astrophysics and Cosmology Group,
        Facultad de Ciencias Astron\'{o}micas y Geofisicas,
        Universidad Nacional de La Plata, La Plata, Argentina\\
        E-mail: \email{guscontrera@gmail.com}}

\author{Rafa{\l} {\L}astowiecki\\
        Institute for Theoretical Physics, University of Wroc{\l}aw,
	Wroc{\l}aw, Poland\\
        E-mail: \email{lastowiecki@ift.uni.wroc.pl}}

\abstract{Nonlocal PNJL models allow for a detailed description of chiral
quark dynamics with running quark masses and wave function renormalization in
accordance with lattice QCD (LQCD) in vacuum.
Their generalization to finite temperature T and chemical potential $\mu$
allows to reproduce the $\mu$-dependence of the pseudocritical temperature from
LQCD when a nonvanishing vector meson coupling is adjusted.
This restricts the region for the critical endpoint in the QCD phase diagram
and stiffens the quark matter equation of state (EoS).
It is demonstrated that the construction of a hybrid EoS for compact star
applications within a two-phase approach employing the nonlocal PNJL
EoS and an advanced hadronic EoS leads to the masquerade problem.
A density dependence of the vector meson coupling is suggested as a possible
solution which can be adjusted in a suitable way to describe hybrid stars with
a maximum mass in excess of $2~M_\odot$ with a possible early onset of quark
deconfinement even in the cores of typical ($M\sim 1.4~M_\odot$) neutron stars.
}

\FullConference{Xth Quark Confinement and the Hadron Spectrum,\\
		October 8-12, 2012\\
		TUM Campus Garching, Munich, Germany}

\begin{document}

\section{Introduction}

The measurement of the high mass of $M=1.97 \pm 0.04~M_\odot$ for the pulsar
PSR J1614-2230 \cite{Demorest:2010bx} has reinforced the question whether the
phase transition to  hyperon matter \cite{Ambartsumyan:1960} or to deconfined 
quark matter \cite{Baym:1976yu} in compact stars can be excluded (see, e.g., 
\cite{Alford:2006vz,Klahn:2006iw,Ozel:2010bz,Klahn:2011fb,
Weissenborn:2011qu,Bednarek:2011gd,Lastowiecki:2011hh,Klahn:2012pi,Sedrakian:2013rr}).

In the present contribution, we want to discuss an answer to this question on 
the basis of a recently developed approach to quark matter within a nonlocal
Nambu--Jona-Lasinio (NJL) model coupled to the Polyakov loop 
\cite{Blaschke:2007np,Contrera:2007wu,Hell:2008cc} 
(denoted here as ``nl-PNJL'') in its extension including wave function 
renormalization (WFR) 
\cite{Contrera:2010kz,Horvatic:2010md,Hell:2011ic}
which goes far beyond the local PNJL models.  

These models have been developed and tested against modern LQCD
data in the finite temperature domain and at small chemical potentials where 
also results from LQCD within Taylor expansion techniques exist 
\cite{Kaczmarek:2011zz}. 
The extension of nl-PNJL models to the full QCD phase diagram 
poses a challenge  \cite{Contrera:2012wj,Hell:2012da}.
In particular their extension to the zero temperature case for 
studies of deconfinement in compact stars has so far been performed without
WFR \cite{Blaschke:2007ri,Orsaria:2012je}. 
The inclusion of the latter into these studies is a novelty on which 
we present first results in this contribution.
  
\section{Nonlocal PNJL model}

The Lagrangian for the $SU(2)_f$ nonlocal models, including vector channel
interactions, is given by
\begin{equation}
{\cal L} = \bar{q}(i\Dslash-m_0) q + {\cal L}_{\rm int}+\mathcal{U}(\Phi)~,
\end{equation}
where $q$ is the $N_{f}=2$ fermion doublet $q\equiv(u,d)^T$,
and $m_0$ is the current quark mass (we consider isospin symmetry, that is
$m_0=m_{u}=m_{d}$). The covariant derivative is defined as
$D_\mu\equiv \partial_\mu - iA_\mu$, where $A_\mu$ are color gauge fields.
The nonlocal interaction channels are given by
 \begin{equation}
{\cal L}_{\rm int}= -\frac{G_{S}}{2} \Big[ j_S(x)j_S(x) + j_P(x)j_P(x)
- j_{p}(x)j_{p}(x)\Big] {-} \frac{G_V}{2} j_V(x)\, j_V(x),
\end{equation}
where the nonlocal currents are
\begin{eqnarray}
j_{a}(x)  &  =&\int d^{4}z\ g(z)\ \bar{q}\left(x+\frac{z}{2}\right)
\ \Gamma_{a}\ q\left(  x-\frac{z}{2}\right)  \ , ~~ a=S,P,V~,\nonumber\\
j_{p}(x)  &  =&\int d^{4}z\ f(z)\ \bar{q}\left(x+\frac{z}{2}\right)
\ \frac{i {\doublearrowslash}}{2\ \kappa_{p}}
\ q\left( x-\frac{z}{2}\right) \ ,
\label{eq:currents}
\end{eqnarray}
with
$\Gamma_{a}=(\Gamma_{S},\Gamma_{P},\Gamma_V)
=(\leavevmode\hbox{\small1\kern-3.8pt\normalsize1},i\gamma_{5}\vec{\tau},\gamma_0)$
for scalar, pseudoscalar and vector currents respectively, and
$u(x^{\prime}){\doublearrowpartial %
}v(x)=u(x^{\prime})\partial_{x}v(x)-\partial_{x^{\prime}}u(x^{\prime})v(x)$.
The functions $g(z)$ and $f(z)$ in (\ref{eq:currents}) are
nonlocal covariant form factors characterizing the corresponding
interactions. The scalar-isoscalar current $j_{S}(x)$
will generate the momentum dependent quark mass in the
quark propagator, while the ``momentum'' current, $j_{p}(x),$ will
be responsible for a momentum dependent WFR of the propagator.
Note that the relative strength between both interaction terms is controlled
by the mass parameter $\kappa_{p}$ introduced in (\ref{eq:currents}).

In what follows it is convenient to Fourier transform into momentum space.
Since we are interested in studying the chiral phase
transition we extend the effective action to finite temperature $T$
and quark chemical potential $\mu$
using the Matsubara formalism.

Concerning the gluon fields we use the same prescription as in previous works
\cite{Contrera:2007wu,Contrera:2010kz}, but in our present case we have chosen
a $\mu$-dependent logarithmic effective potential described in
\cite{Dexheimer:2009va}
\begin{equation}
\mathcal{U}(\Phi,T,\mu)=(a_0T^4+a_1\mu^4+a_2T^2\mu^2)\Phi^2
+ a_3T_0^4\ln{(1-6\Phi^2+8\Phi^3-3\Phi^4)}~,
\label{PL_pot}
\end{equation}
where $a_0=-1.85$, $a_1=-1.44\times 10^{-3}$, $a_2=-0.08$,
$a_3=-0.40$.
In the present work we set the $T_0$ parameter by using the value
corresponding to two flavours $T_0= 208$ MeV, as it has been suggested in
~\cite{Schaefer:2007pw}, and used in subsequent approaches, including the
nonlocal PNJL ~\cite{Pagura:2012} and Polyakov loop-DSE models
~\cite{Horvatic:2010md}.

\begin{center}
\begin{figure}
\includegraphics[scale=0.3]{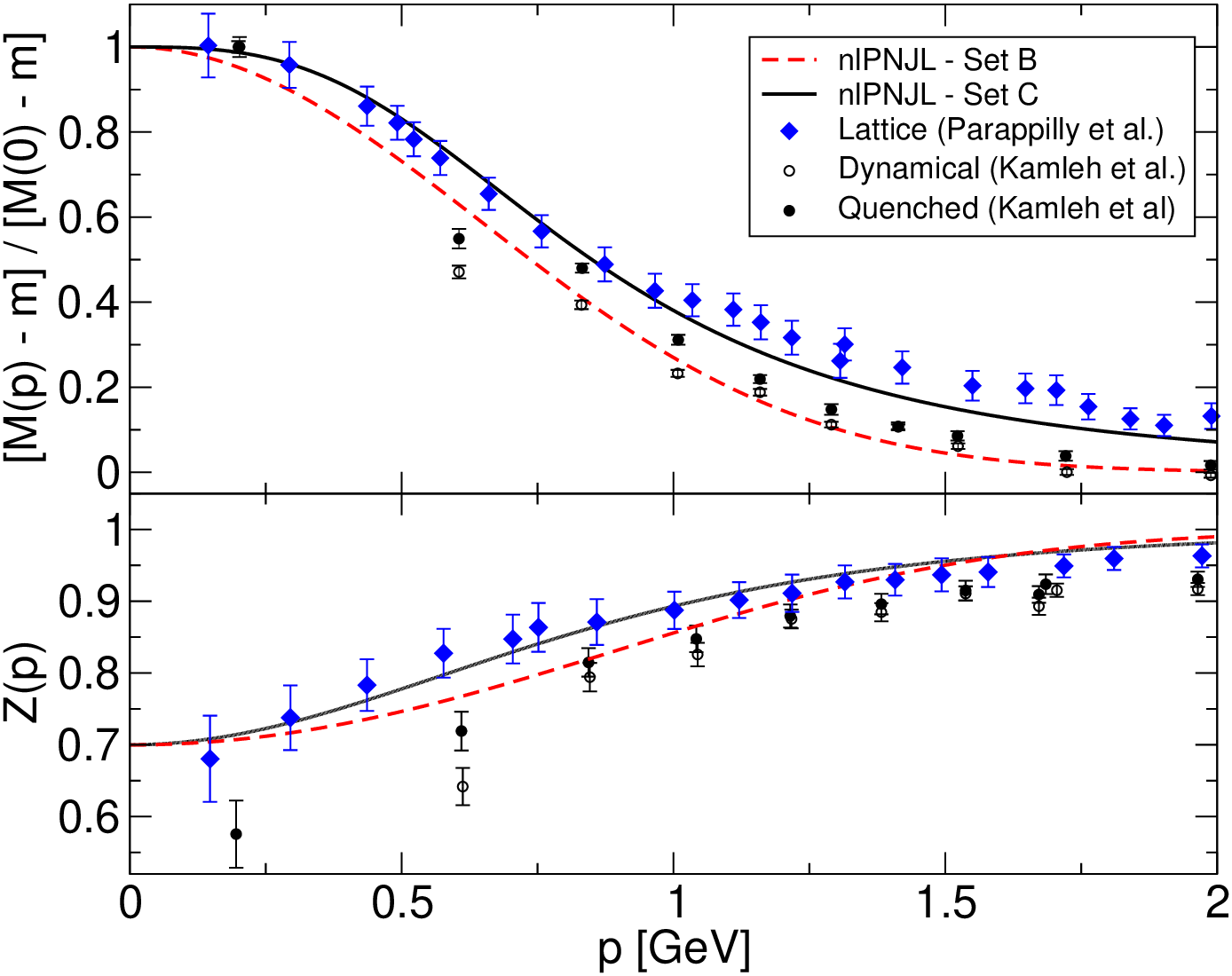}\hspace{5mm}
\includegraphics[scale=0.3]{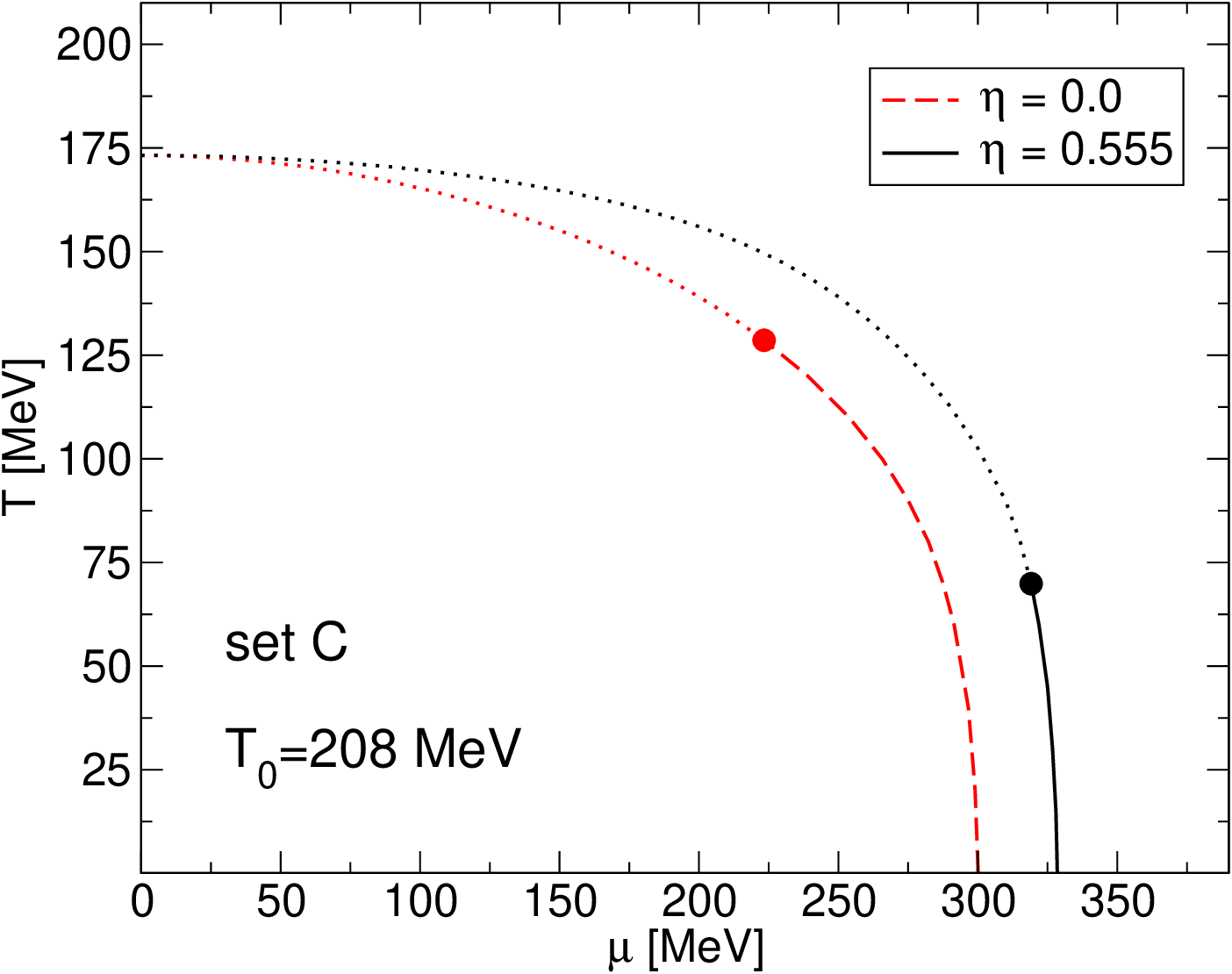}
\caption{(Color online) Left: Normalized dynamical masses and wave function 
renormalization for 
the choice of form factors used in this study (set C), fitted to lattice data 
of Ref.~\cite{Parappilly:2005ei} according to \cite{Noguera:2008cm}.
For comparison, more recent data \cite{Kamleh:2007ud} and alternative 
formfactors according to set B of Ref.~\cite{Contrera:2010kz} are shown.
Right:
The phase diagram of the nl-PNJL in meanfield approximation for the
formfactor parametrization of set C, see also Ref.~\cite{Contrera:2012wj}.
Reproducing the slope of the pseudocritical temperature $\kappa = 0.059$ from
lattice QCD for small chemical potentials \cite{Kaczmarek:2011zz}
requires a scaled vector coupling $\eta=0.555$.
\label{fig1}}
\end{figure}
\end{center}

Finally, to fully specify the nonlocal models under consideration we
fix the model parameters as well as the form factors $g(q)$ and $f(q)$
following \cite{Contrera:2010kz,Noguera:2008cm}, from where we choose the 
more realistic combination of Lorentzian functions (set C, according to the 
notation in Ref.~\cite{Contrera:2010kz}). 
These form factors have been chosen \cite{Noguera:2008cm} such as to 
reproduce the dynamical mass function $M(p)$ and the WFR $Z(p)$ from lattice 
QCD simulations of the quark propagator in the vacuum 
\cite{Parappilly:2005ei}, shown in the left panel of Fig.~\ref{fig1}. 
For comparison, also the more recent data \cite{Kamleh:2007ud} and 
alternative formfactors according to set B of Ref.~\cite{Contrera:2010kz} 
are shown.
Within this framework the mean field thermodynamical potential
$\Omega^{\rm MFA}$ results
\begin{eqnarray}
\Omega^{\rm MFA} =
- {4 T} \sum_{n,c} \int \frac{d^3\vec p}{(2\pi)^3} \, \,
\mbox{ln} \left[ \frac{ (\tilde{\rho}_{n, \vec{p}}^c)^2 +
M^2(\rho_{n,\vec{p}}^c)}{Z^2(\rho_{n, \vec{p}}^c)}\right] 
+\frac{\sigma_1^2 + \kappa_p^2\ \sigma_2^2}{2\,G_S} - \frac{\omega^2}{2 G_V}\
+ \mathcal{U}(\Phi,T)~, 
\label{granp}
\end{eqnarray}
where $M(p)$ and $Z(p)$ are given by
\begin{eqnarray}
M(p) & = & Z(p) \left[m + \sigma_1 \ g(p) \right] ~,~~
Z(p)  =  \left[ 1 - \sigma_2 \ f(p) \right]^{-1}~.
\label{mz}
\end{eqnarray}
In addition, as in \cite{Contrera:2010kz}, we have considered
\begin{equation}
\Big({\rho_{n,\vec{p}}^c} \Big)^2 =
\Big[ (2 n +1 )\pi  T - i \mu + \phi_c \Big]^2 + {\vec{p}}\ \! ^2 \ ,
\label{eq:rho}
\end{equation}
where the quantities $\phi_c$ are given by the relation $\phi =
{\rm diag}(\phi_r,\phi_g,\phi_b)$. Namely, $\phi_c = c \ \phi_3$
with $c = 1,-1,0$ for $r,g,b$, respectively.
In the case of $\Big({\tilde{\rho}_{n,\vec{p}}^c}\Big)$ we have used the same
definition as in (\ref{eq:rho}) but shifting the chemical potential
according to
\begin{equation}
\tilde{\mu} = \mu \; - \omega \; g(p) \; Z(p)~.
\label{mutilde}
\end{equation}

The expression for $\Omega^{\rm MFA}$ as given in (\ref{granp}) turns 
out to be divergent and, thus, needs to be regularized. 
For this purpose we use the same prescription as in
\cite{Contrera:2010kz,GomezDumm:2004sr}.
A necessary condition to find the set of mean field values for 
$\sigma_{1,2}$, $\omega$ and $\phi_3$ 
which would correspond to the absolute minimum of the regularized
thermodynamic potential is the fulfillment of the coupled ``gap'' 
equations  
\begin{equation}
\frac{\partial\Omega^{\rm MFA}_{\rm reg}}{\partial\sigma_{1}} =
\frac{\partial\Omega^{\rm MFA}_{\rm reg}}{\partial\sigma_{2}} =
\frac{\partial\Omega^{\rm MFA}_{\rm reg}}{\partial\omega} =
\frac{\partial\Omega^{\rm MFA}_{\rm reg}}{\partial\phi_3} = 0~.
\label{fullgeq}
\end{equation}
Two of these meanfields are regarded as order parameters as their nonvanishing
value signals the breaking of a symmetry in the system: the quark mass gap 
$\sigma_1$ stands for the breaking of the chiral symmetry and the traced 
Polyakov loop $\Phi=[1+2\cos(\phi_3/T)]/3$ signals the SU(3) center 
symmetry breaking signalling deconfinement.
In the right panel of Fig.~\ref{fig1} we show the phase diagram of the 
chiral transition in the $T-\mu$ plane for two values of the scaled vector
coupling $\eta=G_V/G_S$. 
The dotted lines show the pseudocritical temperatures of the chiral crossover 
transition which at the critical endpoints (CEP), shown as filled circles, go 
over to the critical temperatures (dashed and solid lines) of the first
order transition where the chiral order parameter is discontinuous.    
Within the set C 
parametrization the value $\eta=0.555$ is required for reproducing  
the slope of the pseudocritical temperature $\kappa = 0.059$ 
obtained in LQCD simulations for small chemical potentials 
\cite{Kaczmarek:2011zz}.
In comparison to a vanishing, or lower, vector coupling the CEP is situated at
a lower temperature and higher chemical potential. The CEP position also 
depends on the parametrization of the PNJL model. 
In the local limit it even vanishes, see Ref.~\cite{Contrera:2012wj} for 
details. 

\begin{center}
\begin{figure}
\includegraphics[scale=0.3]{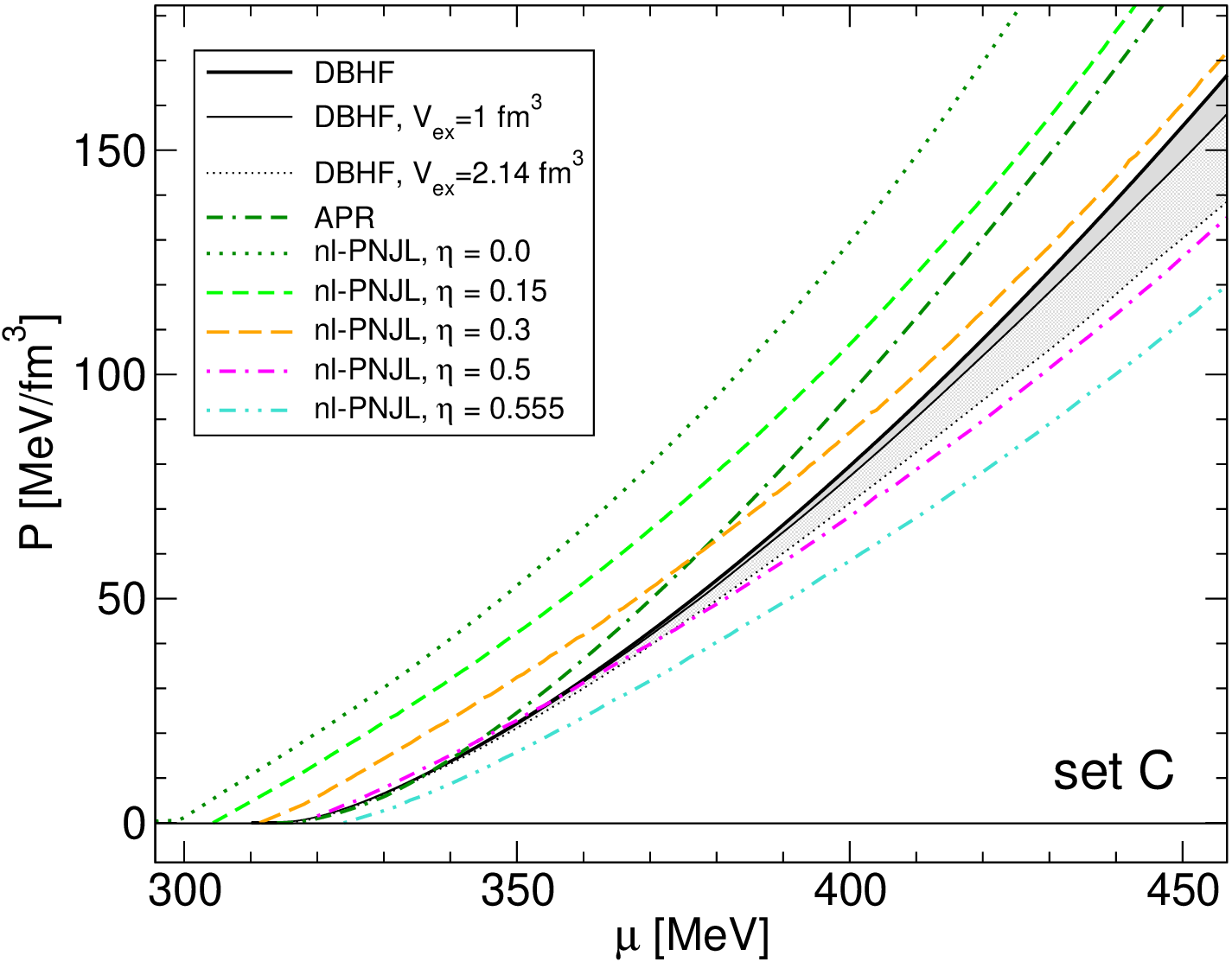}\hspace{5mm}
\includegraphics[scale=0.3]{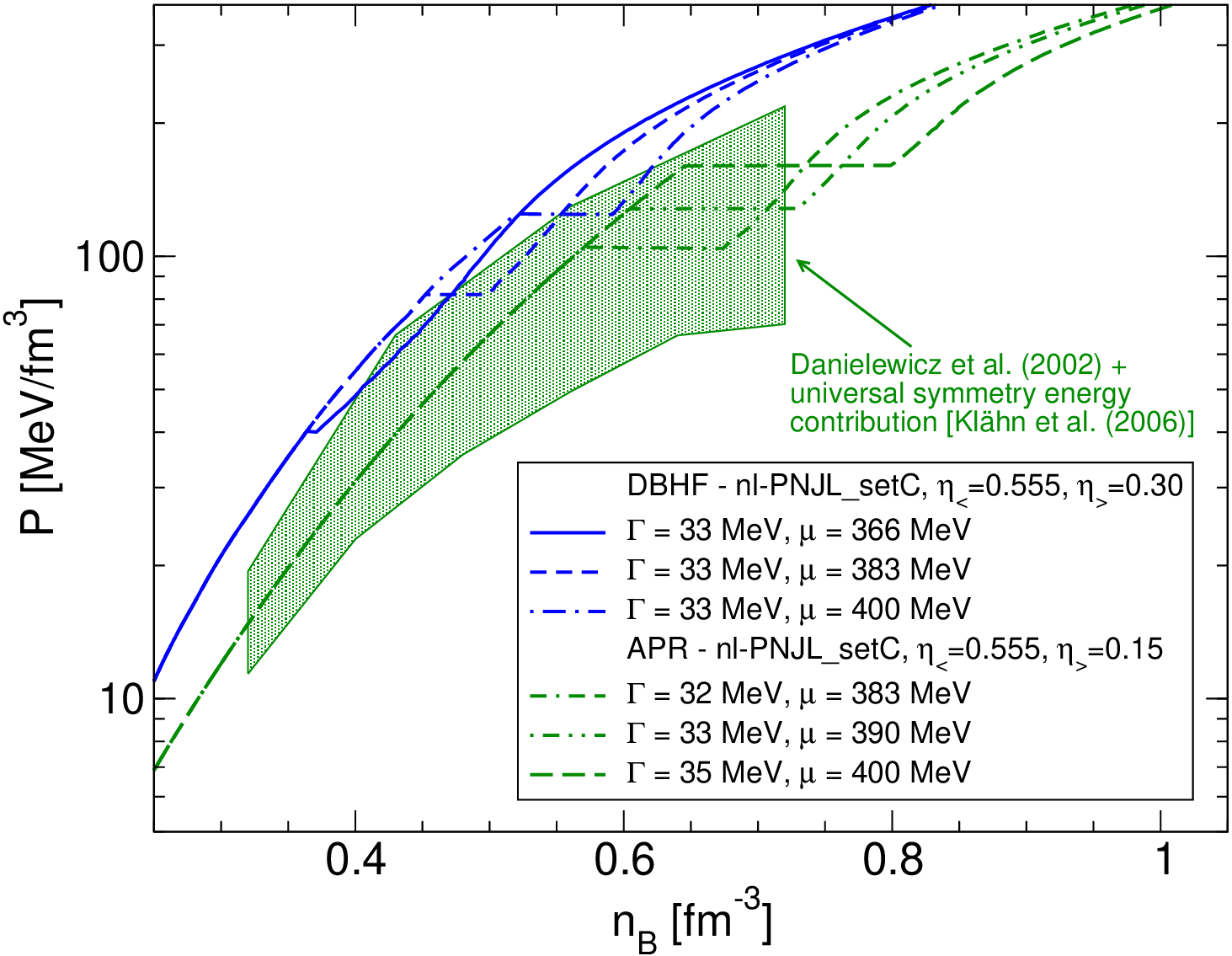}
\caption{(Color online)
Left: 
Pressure vs. baryochemical potential EoS for neutron star matter 
in the hadronic phase (DBHF with and without excluded volume $V_{\rm ex}$, APR)
compared to that of the quark matter phase (nl-PNJL) for different values of 
$\eta$. 
For $\eta\stackrel{>}{\sim} 0.5$ nuclear matter is stable at low 
densities.
For a phase transition to quark matter at high densities (chemical
potentials) the vector coupling has to get reduced to a value 
$\eta_>\stackrel{<}{\sim} 0.3$. 
Right:
Hybrid EoS, where the quark sector is provided using the interpolation 
(\protect\ref{interpol}) and two alternatives for hadronic phase are used
(APR and DBHF).
The phase transition from hadronic to quark matter is determined by 
Eq.~(\protect\ref{Maxwell}).
The green hatched area corresponds to the EoS constraint for symmetric matter
from the flow data analysis \cite{Danielewicz:2002pu} shifted by the 
universal symmetry energy contribution \cite{Klahn:2006ir}.
\label{fig2}}
\end{figure}
\end{center}

The EoS for quark matter is then obtained from 
\begin{equation}
P(T,\mu) = - \min_{\sigma_1,\sigma_2, \omega,\phi_3}\Omega^{\rm MFA}_{\rm reg}
(T,\mu; \{{\sigma}_1, {\sigma}_2, {\omega}, {\phi}_3\})~,
\label{press}
\end{equation}
which holds for a homogeneous system.
On the left panel of Fig.~\ref{fig2} we show the $T=0$ EoS of the nl-PNJL 
model in set C parametrization for different values of $\eta$.

\section{Deconfinement phase transition in compact stars}

The deconfinement phase transition between nuclear matter and quark matter
under compact star conditions\footnote{Here we consider local charge 
neutrality and $\beta-$ equilibrium. A more elaborated treatment considers 
finite size structures (``pasta'') in the mixture of hadronic and quark matter.
See  \cite{Yasutake:2012dw} for a recent discussion and references.} 
is constructed by solving the corresponding condition for phase equilibrium 
\begin{equation}
\label{Maxwell}
P_Q(\mu_c)=P_H(\mu_c)~, 
\end{equation}
where $\mu_c$ is the critical chemical potential for the transition. 
The index $H$ stands for ``hadronic'' and here we will use one of the standard 
nuclear matter EoS: APR \cite{Akmal:1998cf} or DBHF \cite{vanDalen:2004pn}.
The index $Q$ stands for ``quark'' and the corresponding EoS will be introduced
based on the following discussion.
Inspecting the left panel of Fig.~\ref{fig2}, we observe that choosing for 
$P_Q({\mu})=P(0,{\mu})$, i.e. the EoS (\ref{press}) for the nl-PNJL with a 
constant vector coupling strength $\eta$ there occurs one of the following 
problems:
(a) quark matter is favorable for all chemical potentials, even at lowest 
densities (for $\eta\stackrel{<}{\sim} 0.3$);
(b) nuclear matter is favorable for all chemical potentials, even at highest
densities (for $\eta\stackrel{>}{\sim} 0.5$); 
(c) quark and hadronic EoS cross each other either in the ``wrong'' way 
(confinement at high densities) or more than once 
(for $\eta\sim 0.4 \dots 0.5$) since they are marginally indistinguishable 
in the relevant range of chemical potentials.
The latter situation has been called the ``masquerade'' problem
\cite{Alford:2004pf}. 
    
In this situation we suggest the following procedure to avoid unphysical 
solutions and the masquerade problem. 
We anticipate a chemical potential dependence of the scaled vector coupling, 
$\eta(\mu)$, in such a way that at low chemical potentials the value 
$\eta_<=0.555$ is set which is necessary to reproduce the slope of the 
pseudocritical temperature and at high chemical potentials, beyond a fiducial 
range for the deconfinement transition, an asymptotic value is attained which
makes the quark matter EoS favorable over the hadronic one under consideration.
For $H=$ DBHF, this is $\eta_>\le 0.3$.    
For the transition between both regions we suggest here an interpolation ansatz
\begin{eqnarray}
\label{interpol}
P_Q(\mu)&=&P(0,\mu;\eta_<)f_{<}(\mu)+P(0,\mu;\eta_>) f_{>}(\mu)~,\\
f_{\stackrel{<}{>}}(\mu)&=& \frac{1}{2}\left[1 \mp 
		\tanh \left(\frac{\mu-\bar{\mu}}{\Gamma}\right)\right]~. 
\end{eqnarray}
For the two free parameters of the interpolation we suggest the ranges
$\bar{\mu} \sim 360 \dots 410 $ MeV and $\Gamma \sim 30$ MeV.   
The motivation of this choice comes from considering the influence of a 
nucleonic excluded volume, shown also in the left panel of Fig.~\ref{fig2},
which stems from the quark Pauli blocking between nucleons and thus estimates 
the onset of quark substructure effects like the overlap of hadronic wave 
functions in dense matter modifying effective quark-meson couplings. 

Two sets of resulting hybrid EoS, based on DBHF and APR as 
hadronic EoS, respectively, are obtained by the Maxwell construction 
according to Eq.~(\ref{Maxwell}).
They are shown on the right panel of Fig.~\ref{fig2} together with a constraint
for compact star EoS obtained from the flow data analysis for symmetric
matter \cite{Danielewicz:2002pu} shifted by a contribution due to the 
universal symmetry energy of Ref.~\cite{Klahn:2006ir}.

\begin{center}
\begin{figure}
\begin{centering}
\includegraphics[width=0.78\textwidth,height=0.52\textwidth]{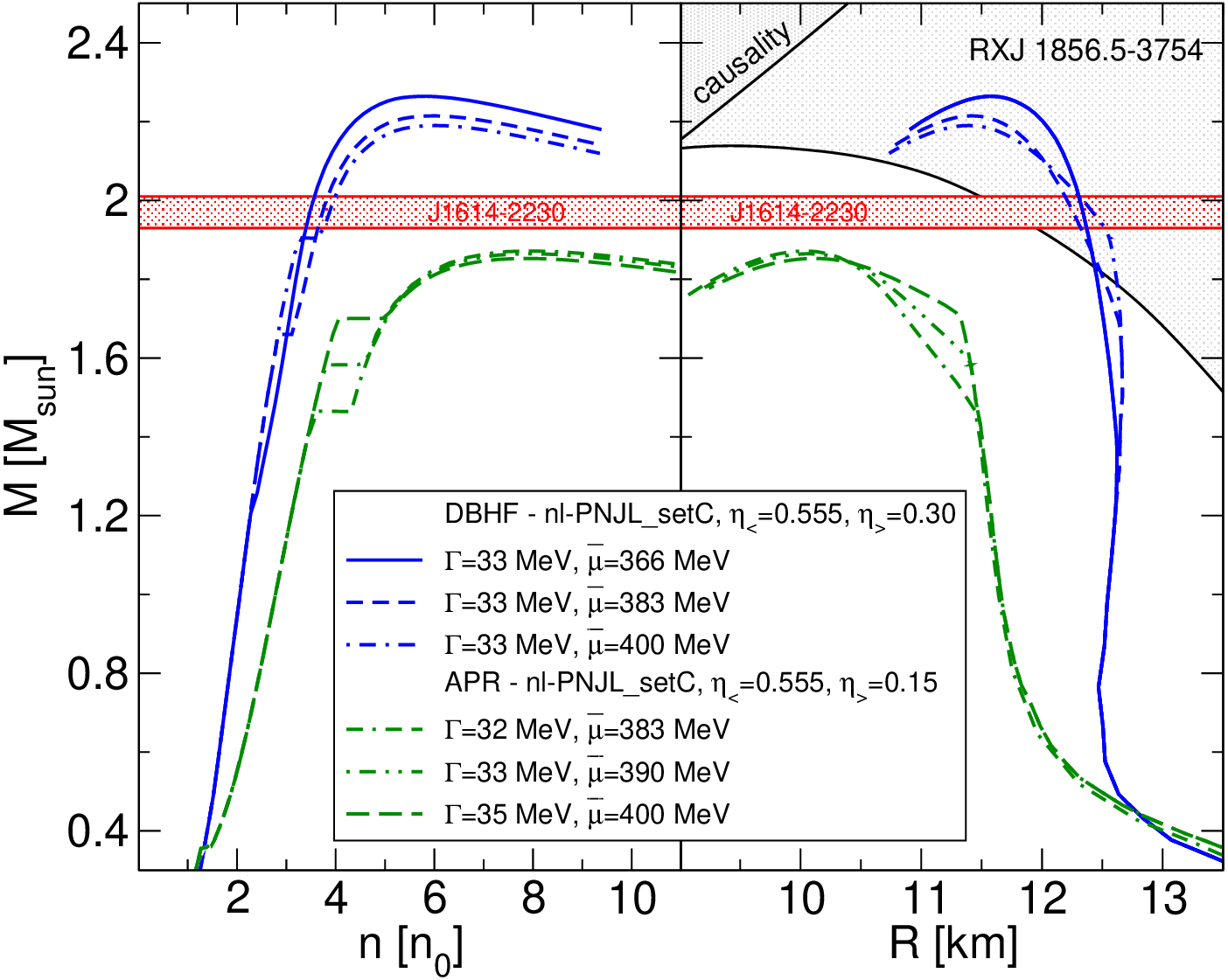}
\end{centering}
\caption{(Color online) Hybrid star sequences based on the DBHF and APR 
EoS for nuclear 
matter and the nl-PNJL model for quark matter proposed here, shown 
together with constraints on the maximum mass from PSR J1614-2230 
\cite{Demorest:2010bx} and on the mass-radius relation from RX J1856.5-3754 
\cite{Trumper:2003we}.}
\label{fig3}
\end{figure}
\end{center}

With these hybrid EoS the Tolman--Oppenheimer--Volkoff equations 
\cite{Tolman:1939jz,Oppenheimer:1939ne} 
have been solved and the resulting sequences of nonrotating 
compact star configurations have been obtained. 
The corresponding dependencies of the mass $M$ of the stars on their radius 
$R$ and on their central density $n(0)$ are shown the left and right panels 
of Fig.~\ref{fig3}, respectively.  

The hybrid star sequences based on the APR EoS do not fulfill the constraint 
on the maximum mass from the mass measured for PSR J1614-2230 
\cite{Demorest:2010bx} and also do not obey the constraint on the minimal 
radius for a given mass from RX J1856.5-3754 \cite{Trumper:2003we} 
since these hybrid EoS describe too ``soft'' compact star matter
\footnote{Note that the analysis of X-ray burst sources in 
Ref.~\cite{Steiner:2012xt} can provide at best a lower limit on the star 
radius which is embedded in that for RX J1856.5-3754, 
see Ref.~\cite{Trumper:2011} for a discussion.}.
The hybrid stars based on the DBHF EoS fulfill both constraints and,
depending on the specific $\mu-$ dependence of the vector coupling strength
$\eta(\mu)$, can have an onset of quark matter in their interiors already at 
a mass as low as $1.2~M_\odot$. 
This would include all compact stars with measured masses known so far. 
However, these EoS are too stiff for the constraint from 
flow data \cite{Danielewicz:2002pu} combined with the universal symmetry 
energy contribution \cite{Klahn:2006ir}. 

\section{Results and Conclusions}

A nonlocal Polyakov-NJL model with vector meson meanfield has been calibrated 
with lattice QCD data for the momentum dependence of the wave function 
renormalization and dynamical mass of the quark propagator in vacuum as well 
as for the slope of the pseudocritical temperature at small chemical potentials
which fixes the vector coupling strength under these conditions.

A straightforward application to nonzero temperatures and chemical potentials
within the Matsubara formalism results in a prediction for the position of the
critical endpoint. 

Considering compact star matter at $T=0$ results in the conclusion that 
quark matter in compact stars cannot occur since its pressure stays always 
below that of standard hadronic EoS for all relevant baryon densities, unless
a medium dependence of basic nl-PNJL model parameters is invoked.
A generic ansatz for the chemical potential dependence of the vector coupling 
strength is presented which allows to estimate the ballpark for such a 
dependence allowing hybrid star configurations which fulfill basic compact 
star constraints. Examples illustrate the sensitivity to the nuclear EoS:
While the set of hybrid star EoS based on APR is too soft to fulfill mass and 
radius constraints, 
those based on DBHF 
satisfy these constraints but 
appear too stiff for the flow data from heavy-ion collisions. 
Thus compact star observations and heavy-ion collisions together may guide the 
path towards the EoS of dense QCD matter.

\section*{Acknowledgement}
The authors acknowledge support from CompStar, a Research Networking Programme
of the European Science Foundation and the programme of the Polish
Ministry for Science and Higher Education for supporting it.
D.B. profited from inspiring discussions with Tetsuo Hatsuda, Ken-Ichi Imai, 
Toshiki Maruyama, Kota Masuda, Thomas Rijken, Tatsuyuki Takatsuka, Nobutoshi 
Yasutake 
at the J-PARC workshop and the SNP school during his visit at the Research 
Group for Hadron Physics at the Advanced Sciene Research Center of JAEA in 
Tokai where this work was completed.
He is grateful for financial support from RFFI under grant number 11-02-01538a
and the JAEA Foreign Researcher Inviting Program. 
D.B. and D.E.A.C. were supported by the Polish National Science Center under 
the ``Maestro'' programme.
S.B. acknowledges support of the Croatian Ministry of Science, Education
and Sports under contract No. 119-0982930-1016.
R.{\L}. received support from the faculty of Physics and Astronomy
at the University of Wroc{\l}aw under grant number 2291/M/IFT/12.

\end{document}